\documentclass[aps,eqsecnum]{revtex4}
\usepackage{epsfig}
\usepackage{graphicx}
\usepackage{amsmath}
\usepackage{amssymb}
\usepackage{color}
\newcommand{\e}{\epsilon}
\newif\ifold             \oldtrue            
\def\ba{\begin{eqnarray}}
\def\ea{\end{eqnarray}}

\def\bp{\mathbf{p}}
\def\bk{\mathbf{k}}
\def\bq{\mathbf{q}}

\newcommand{\be}{\begin{equation}}
\newcommand{\ee}{\end{equation}}

\newcommand{\A}{\alpha}
\newcommand{\pr}{\partial}

\begin{document}
\title{Supercritical Coulomb center and excitonic instability  in graphene}
\author{O.~V.~Gamayun,\, E.~V.~Gorbar,\, and V.~P.~Gusynin}
\email{gamayun@bitp.kiev.ua, gorbar@bitp.kiev.ua, vgusynin@bitp.kiev.ua}
\affiliation{Bogolyubov Institute for Theoretical Physics, 14-b Metrologichna
str., Kiev 03680, Ukraine}

\begin{abstract}
It is well known that there are resonant states with complex energy for the
supercritical Coulomb impurity in graphene. We show that opening of a
quasiparticle gap decreases the imaginary part of energy, $|{\rm Im}E|$, of
these states and stabilizes the system. For gapless quasiparticles with strong
Coulomb interaction in graphene, we solve the Bethe-Salpeter equation for the
electron - hole bound state and show that it has a tachyonic solution for
strong enough coupling $\alpha=e^{2}/\kappa\hbar v_{F}$ leading to instability
of the system. In the random phase approximation, the critical coupling is
estimated to be $\alpha_{c} =1.62$ and is an analogue of the critical charge in the
Coulomb center problem. We argue that the excitonic instability should be
resolved through the formation of an excitonic condensate and gap generation in
the quasiparticle spectrum.

\end{abstract}

 \maketitle

\section{Introduction}

Graphene is an one-atom-thick layer of graphite packed in the honeycomb
lattice. Although theoretically considered long time ago \cite{gr}, graphene
became an active area of research only recently after the experimental
fabrication \cite{novos} of this material and because of a variety of
its unusual electronic properties.

At low energy the band structure of graphene is formed by the $\pi$-electron
orbits of carbon and consists of the valence and conduction bands and
corresponds to cones touching each other at the so-called Dirac points.
Quasiparticle excitations close to these points are described by the massless
Dirac equation and have a relativistic-like dispersion $E=\pm \hbar v_F|\mathbf{k}|$,
where $v_{F}\approx 10^{6}m/s$ is the Fermi velocity and $\mathbf{k}$ is
the quasiparticle wave vector. This fact brings an exciting connection between
graphene and $3+1$- dimensional quantum electrodynamics (QED).

The vanishing density of states at the Dirac points ensures that the
Coulomb interaction between the electrons in graphene retains its
long-range character in view of vanishing of the static polarization
function for $q \to 0$ \cite{Gonzalez}. The large value of the
coupling constant $\alpha=e^2/\hbar v_F \sim 1$ means that a strong
attraction takes place between electrons and holes in graphene and
this resembles strongly coupled QED, thus providing an opportunity
for studying the strong coupling phase experimentally within
a condensed matter laboratory. Given the strong attraction, one may
expect an instability in the excitonic channel in graphene with
subsequent quantum phase transition to a phase with gapped
quasiparticles that may turn graphene into an insulator. This
semimetal-insulator transition in graphene is widely discussed now
in the literature \cite{Drut,Dillenschneider} since the first study
of the problem in Refs.\cite{Khveshchenko,Gra2002}. The gap opening
is similar to the chiral symmetry breaking phenomenon
that occurs in strongly coupled QED and was studied in the
70-ties and 80-ties
\cite{Fomin,review,Bardeen,Miranskybook,Kogut}. In fact, the
predicted strong coupling phase of QED,  like other QED effects not
yet observed in nature (Klein tunneling, Schwinger effect, etc.), has
a chance to be tested in graphene.

We begin our study  with the problem of the supercritical Coulomb
center in Sec.~\ref{Coulomb-center} in graphene. As is well known \cite{Khalilov,Novikov},
for the Coulomb potential, $V_{C}(r)=-Ze^{2}/\kappa r$, the spectrum of
quasiparticles with a gap $\Delta$ contains a continuum spectrum for
$|E|>\Delta$ and a discrete one for $0<E<\Delta$. The lowest bound state energy
equals $E_{0}=\Delta \sqrt{1-(2Z\alpha)^{2}}$ and becomes purely imaginary for
$Z\alpha>1/2$ - the ``fall into the center'' phenomenon. The unphysical complex
energies indicate that the Hamiltonian of the system is not a self-adjoint
operator for supercritical values $Z\alpha>1/2$ and should be extended to
become a self-adjoint operator. The way out of this situation is well known
from the study of the Dirac equation in QED: one should replace the singular
1/r potential by a regularized potential which takes into account the finite
size of the nucleus, $R$, \cite{Pomeranchuk,Popov-center,Popov}. When the
charge Z increases, the energies of discrete states approach the negative
energy continuum, $E=-\Delta$, and then dive into it. Then discrete states
turn into resonances with a finite lifetime, which can be described as
quasistationary states with complex energies, ${\rm Im}E\neq0$. Such states
correspond to a rearrangement process when an electron-hole pair is created
from the vacuum, the positively charged hole goes to infinity and the electron
is coupled to the Coulomb center, thus shielding the charge of the latter. The
critical charge $Z_{c}$ is determined by the condition of appearance of nonzero
imaginary part of the energy and
increases with the increase of $\Delta$.

Turning to the case of gapless quasiparticles in the regularized
Coulomb potential, there are no
discrete levels for $Z\alpha<1/2$ due to  scale
invariance of the massless Dirac equation, and for $Z\alpha>1/2$
quasistationary states emerge \cite{Shytov1}. The energy
of quasistationary levels for the regularized potential has a characteristic
exponential-type dependence, ${\rm Re}E,\,\, {\rm Im}E\sim
-R^{-1}\exp(-\pi/\sqrt{(Z\alpha)^{2}-1/4})$, in the nearcritical regime
(according to the analysis in Appendix A, the critical coupling $Z_c\alpha \to
1/2$ for $R\Delta \to 0$). We find that  switching on a fermion gap,
$\Delta\ll|E|$, decreases $|{\rm Im}E|$, i.e. increases the stability of the
system. The situation here is analogous to the problem of a massless electron
in the supercritical Coulomb center in QED first studied in
\cite{FominMiransky} (for a review, see, \cite{review}).

In Sec.\ref{excitonic-instability} we show that the instability in the
supercritical Coulomb center problem is closely related to the
excitonic instability in graphene in the supercritical coupling
constant regime $\alpha>\alpha_{c}\sim1$.
Solving the Bethe--Salpeter equation for an electron - hole bound state in graphene, we
demonstrate that for strong enough coupling constant there are tachyon states with
imaginary energy ($E^{2}<0$) in the spectrum which play here the role of the quasistationary states
in the problem of the supercritical Coulomb center. The presence of tachyons signals
 that the normal state of freely standing graphene is unstable. In fact,
the tachyon instability can be viewed as the field theory analogue of the ``fall into the center''
phenomenon and the critical coupling $\alpha_{c}$ is an analogue of the critical coupling
constant $Z_{c}e^{2}/\hbar v_{F}$ in the problem of the Coulomb center.
However, in view of the many-body character of the problem, the way of curing
the instability in graphene (like in QED \cite{review}) is quite
different from that in the case of the
supercritical Coulomb center. Since the coupling constant
in freely standing graphene $\alpha\approx2.19$ is larger than ${1}/{2}$, the
quasielectron in graphene has the supercritical Coulomb charge.  This leads to
the production of an electron-hole pair, the hole is coupled to the initial
quasielectron  forming a bound state but the emitted  quasielectron  has again a supercritical
charge. Thus the process of creating pairs
continues leading to the formation of exciton (chiral) condensate in the stable phase,
and, as a result, the quasiparticles acquire a gap.
The exciton condensate formation resolves the problem of instability, hence
 a gap generation should take place in a freely standing graphene making it an insulator.

 Sec.IV contains our conclusions. In Appendix \ref{discretespectrum}
we consider the behavior of bound states for gapped graphene quasiparticles
in the regularized Coulomb center and find the critical coupling $Z_{c}\alpha$
as a function of the parameter $R\Delta$ for the lowest energy level.
In Appendix \ref{Pochhammer:eq} we give the exact solution for the tachyon wave function
which satisfies the fourth order differential equation.

\section{ The supercritical Coulomb center: resonant states}
\label{Coulomb-center}

Although the electron-hole problem is a many body problem in graphene, it is
instructive to consider a rather simple one particle problem of the electron in the
field of the supercritical Coulomb center in view of the connection of the
latter problem with the excitonic instability in graphene (for a similar
problem of instability in the case of a massless fermion in the external field
of the supercritical Coulomb center in QED see
Refs.\cite{FominMiransky,review}).

The Coulomb center problem was studied quite in detail in the literature
\cite{Khalilov,Novikov,Shytov1,Fogler,Pereira,Terekhov,Shytov2,Castro}. Since
we are mainly interested here in resonant states, we will consider a
regularized Coulomb potential
\be
V(r) = -\frac{Ze^{2}}{\kappa r},
\,\,(r>R),\quad V(r) = -\frac{Ze^{2}}{\kappa R}, \,\, (r<R),
\ee
($\kappa$ is a dielectric constant) because resonant states are connected with diving into the
lower continuum that takes place only in the case of a regularized potential
\cite{Popov, Greiner}.

The electron quasiparticle states in vicinity of the $K$ point  of graphene in
the field of Coulomb impurity are described by the Dirac Hamiltonian in 2+1
dimensions
\be
\mathcal{H}=\left (\sigma^3 \Delta +V(r) -i\hbar v_F \sigma^1
\pr_x - i\hbar v_F \sigma^2 \pr_y \right ), \label{Hamiltonian}
\ee
where
$\sigma^i$ are Pauli matrices, $v_F$ the Fermi velocity. (The Hamiltonian of
quasiparticle excitations near the $K^{\prime}$ point is given by
(\ref{Hamiltonian}) with matrices $\sigma^i$  multiplied by -1.) Note that we
introduced the Dirac gap $\Delta$. Although it is absent in the quasiparticle
Hamiltonian in graphene in view of the $U(4)$ symmetry, it may appear due to
spontaneous symmetry breaking. Since Hamiltonian (\ref{Hamiltonian}) commutes
with the total angular momentum operator $J_z=L_z+S_z=-i\hbar
\frac{\partial}{\partial \phi}+\frac{\hbar}{2}\sigma_3$, we seek eigenfunctions
in the following form:
\be \Psi = \frac{1}{r}\left(\begin{array}{c}
          e^{i\phi (j-1/2)}\,a(r) \\
          i\,e^{i\phi (j+1/2)}\,b(r)
        \end{array}\right).
\label{spinor}
\ee
Then we obtain a system of two coupled ordinary differential
equations of the first order
 \be
a' - (j+1/2)\frac{a}{r} + \frac{E+\Delta-V(r)}{\hbar v_{F}}b =0,
\quad\quad\quad\quad\quad
 b' + (j-1/2)\frac{b}{r} - \frac{E-\Delta-V(r)}{\hbar v_{F}}a =0.
 \quad\quad\quad\quad
\label{1} \ee
It is convenient to define the variables $\e=
{E}/{\hbar v_F},\, m= {\Delta}/{\hbar v_F}$,
$u=\sqrt{m^{2}-\epsilon^{2}}$, $\rho=2u r$, and $\alpha
={e^2}/{\hbar v_F\kappa}$. Equations (\ref{1}) are solved in Appendix
\ref{discretespectrum} where the discrete spectrum is found in the weak
coupling regime $Z < Z_c$. According to the analysis there, the critical
coupling $Z_c\alpha \to 1/2$ for $mR \to 0$.

Let us analyze Eqs.(\ref{1}) in the supercritical case $Z\alpha >
1/2$ and show that there are resonant states for $|\epsilon|
> m$ (we define the gap $\Delta>0$). These states describe the instability of
the supercritical charge problem with respect
to the creation of electron-hole pairs from the vacuum. The created electron is coupled
to the Coulomb center, thus shielding the charge of the latter while positively
charged hole goes to infinity \cite{Popov,Greiner}; the process is repeated until
the charge of the Coulomb center is reduced to a subcritical value.

The Whittaker function $W_{\mu,\nu}(\rho)$ with $\mu=1/2
+Z\alpha\epsilon/u$, $\nu=\sqrt{j^{2}-Z^2\alpha^{2}}$ describes bound
states for $|\epsilon|<|m|$ which are situated on the first physical sheet of
the variable $u$ and for which ${\rm Re}u>0$ (see,
Eq.(\ref{boundstate-wf})). The quasistationary states are described by the
same function $W_{\mu,\nu}(\rho)$ and are on the second unphysical sheet with
${\rm Re}u<0$. We shall look for the solutions corresponding to the
quasistationary states which define outgoing hole waves at $r\to\infty$ with
\be
{\rm Re}\epsilon<0,\quad {\rm Im}\epsilon<0, \quad {\rm
Re}u<0,\quad {\rm Im}u<0.
\ee
For solutions with $Z^2\alpha^2 > j^2$ resonance states are determined by
Eq.(\ref{matching}) for bound states
where $\nu$ is replaced by $\nu=i\beta$. We will consider the states with
$j=1/2$ which correspond to the $nS_{\frac{1}{2}}$-states, in particular, the
lowest energy state belongs to them. The corresponding equation then takes the
form
\be
\frac{W_{\frac{1}{2}+\frac{Z\alpha\epsilon}{u},i\beta}(\rho)}{\left(\frac{1}{2}
-\frac{Z\alpha
m}{u}\right)W_{-\frac{1}{2}+\frac{Z\alpha\epsilon}{u},i\beta}
(\rho)}\Big|_{r=R} = \frac{k+1}{k-1},\,\,\,\, k =
\,\frac{m+\e}{u}\sqrt{\frac{\e+Z\alpha/R-m}{\e+Z\alpha/R+m}}
\frac{J_{1}(\widetilde{\rho})}{J_{0}(\widetilde{\rho})},\,
\widetilde{\rho}=\sqrt{(Z\A+\e R)^2-m^2R^2},
\ee
where $W_{\mu,\nu}(x)$ and $J_{a}(x)$ are the Whittaker and Bessel functions, respectively.

We are interested in the case of $|\epsilon|\gg m$, and more important, in the
case of the massless electron, $m=0$. The analytical results can be obtained
for the near-critical values of $Z$ when $Z\alpha-1/2 \ll 1$. We assume that
$|2u R| \ll 1$, then using the asymptotic of the Whittaker function, we
find
\be
(2u R)^{2i\beta}\frac{\Gamma(1-2i\beta)}{\Gamma(1+2i\beta)}\frac{\Gamma\left(1+i\beta
-\frac{Z\alpha\epsilon}{u}\right)}
{\Gamma\left(1-i\beta-\frac{Z\alpha\epsilon}{u}\right)}=
\frac{\frac{1}{2}-i\beta -
\frac{Z\alpha(m-\epsilon)}{u}}{\frac{1}{2}+i\beta
-\frac{Z\alpha(m-\epsilon)}{u}}
\frac{\frac{1}{2}+i\beta-Z\alpha\frac{J_{1}(Z\A)}{J_{0}(Z\A)}}
{\frac{1}{2}-i\beta-Z\A\frac{J_{1}(Z\A)}{J_{0}(Z\A)}}\,.
\label{matching-resonance-final}
\ee
Expanding Eq.(\ref{matching-resonance-final}) in the near critical region in powers of
$\beta=\sqrt{Z^2\alpha^2-1/4}$, we find the following equation:
\be
(-2i\sqrt{\e^2-m^2} R)^{2i\beta} =
1+4i\beta\left[\frac{J_0(1/2)}{J_0(1/2)-J_1(1/2)}+\Psi(1)-\frac{1}{2}
\Psi\left(1-\frac{i}{2}\frac{\e}{\sqrt{\e^2-m^2}}\right)-
\frac{1}{1+i\sqrt{\frac{\e-m}{\e+m}}}\right].
\label{resonance-first-expansion}
\ee
Here $\Psi(x)$ is the psi-function and we
put $u=-i\sqrt{\epsilon^{2}-m^{2}}$ where ${\rm
Im}\sqrt{\epsilon^{2}-m^{2}}<0$ on the second sheet. At first we consider the
case $m=0$. Writing $\epsilon=|\epsilon|e^{i\gamma}$
Eq.(\ref{resonance-first-expansion}) takes the form
\be
\ln(2|\epsilon|R)+i\left(\gamma-\frac{\pi}{2}\right)\simeq2\left[\frac{J_0(1/2)}
{J_0(1/2)-J_1(1/2)}+\Psi(1)-\frac{1}{2}\Psi\left(1-\frac{i}{2}\right)-\frac{1}{1+i}\right]
-\frac{\pi n}{\beta}, \quad n=1,2,\dots.
\ee
We find
\be
\epsilon^{(0)}_{n}=aR^{-1}e^{i\gamma}\exp\left[-\frac{\pi
n}{\sqrt{Z^2\alpha^{2}-1/4}}\right]= -(1.18+0.17i)R^{-1}\exp\left[-\frac{\pi
n}{\sqrt{Z^2\alpha^{2}-1/4}}\right], \quad n=1,2,\dots,
\label{resonant-states}
\ee
where
\ba
&&\gamma=\frac{\pi}{2}\left(1+\coth\frac{\pi}{2}\right)\approx3.28,\\
&&a=\frac{1}{2}\exp\left[\frac{2J_0(1/2)}{J_0(1/2)-J_1(1/2)}+2\Psi(1)-1-{\rm
Re}\Psi\left(1 -\frac{i}{2}\right)\right]\approx1.19.
\ea
The energy of quasistationary states (\ref{resonant-states}) has a characteristic
essential-singularity type dependence on the coupling constant reflecting the
scale invariance of the Coulomb potential. The infinite number of
quasistationary levels is
related to the long-range character of the Coulomb potential.
Note that a similar dependence takes place in the supercritical Coulomb center problem in
QED \cite{FominMiransky}. Our results are also in agreement with
Ref.\cite{Shytov1}.

Since  the ``fine structure constant''
$e^{2}/\hbar v_{F}\approx2.19$ in  graphene, an instability appears already for the charge
$Z=1$. However, in the analysis above we did not take into account
the vacuum polarization effects. Considering these effects and treating the
electron-electron interaction in the Hartree approximation, it was shown in
Ref.\cite{Terekhov} that the effective charge of impurity $Z_{eff}$ is such
that the impurity with bare charge $Z=1$ remains subcritical,
$Z_{eff}e^{2}/(\kappa\hbar v_{F})<1/2$, for any coupling $e^{2}/(\kappa\hbar v_{F})$, while
impurities with higher $Z$ may become supercritical.

For finite $m$ and in the case $|\e| \gg m,\,\,\mbox{Re}\,\e < 0$, expanding
Eq.(\ref{resonance-first-expansion}) in $m/\epsilon$ we get up to the terms of order
$m^2/\epsilon^2$,
\be
\e-\frac{m^2}{2\e} = \e^{(0)}_n
\left(1-\frac{m}{\e}+\frac{m^2}{\e^2}(0.29-0.23i)\right),\quad\quad\quad
n=1,2,\dots\, .
\label{energy-level}
\ee
The resonant states with
$\e^{(0)}_n$ describe the spontaneous emission of positively charged holes when
electron bound states dive into the lower continuum in the case $m=0$. In order
to find corrections to these energy levels due to nonzero $m$, we seek solution
of Eq.(\ref{energy-level}) as a series $\e= \sum\limits_{k=0}^{\infty}\e^{(k)}$
with $\e^{(k)}$ of order $m^k$ and easily find the first two terms
\be
\epsilon_{n}=\epsilon_{n}^{(0)}-m+ \frac{m^2}{|\e^{(0)}_n|}(0.24+0.20i).
\ee
Since ${\rm Im}\epsilon_{n}^{(0)}<0$, the appearance of a gap results in
decreasing the width of resonance and, therefore, increases stability of the
system.

We considered above the case $|\epsilon| \gg m$ and analyzed how a nonzero mass
affects resonant states. It is instructive to consider resonant states also in
the vicinity of the level $\epsilon = - m$ when bound states dive into the
lower continuum and determine their real and imaginary parts of energy. First
of all, nonzero $m$ increases the value of the critical charge. Let us find it.
Using Eq.(\ref{critcoupling:eq}) in Appendix A, we obtain that the critical
value $Z_c\alpha$ for $j=1/2$ scales with $m$ like (see Fig.1)
\be
Z_{c}\alpha\simeq \frac{1}{2}+\frac{\pi^{2}}{\log^{2}(cmR)}, \quad
c=\exp\left[-2\Psi(1)-\frac{2J_{0}(1/2)}{J_{0}(1/2)-J_1(1/2)} \right] \approx
0.21.
\ee
Note that the dependence of the critical coupling on $mR$ is quite
similar to that in the strongly coupled QED \cite{Fomin,review}.
\begin{figure}[h]
\includegraphics[width=6.5cm]{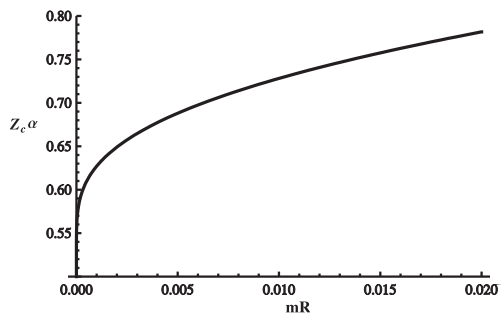}
\caption{The critical coupling as a function of $mR$ for the $1S_{1/2}$ level.}
\label{c1}
\end{figure}
For $Z>Z_c$, using Eq.(\ref{resonance-first-expansion}), we find the following
resonant states:
\begin{equation}
\epsilon=-m\left(1+\xi+i\frac{3\pi}{8}e^{-\pi/\sqrt{2\xi}}\right),\quad
\xi=\frac{3\pi}{8}\,\frac{\beta-\beta_c}{\beta\beta_c},
\label{massive-resonant-states}
\end{equation}
where $\beta_c=\sqrt{(Z_c\alpha)^{2}-1/4}$. Like in QED \cite{Popov-center} the
imaginary part of energy of these resonant states vanishes exponentially as $Z
\to Z_c$. Such a behavior is connected with tunneling through the Coulomb
barrier in the problem under consideration. For the quasielectron in graphene
in a central potential $V(r)$, expressing the lower component of the
Dirac spinor (\ref{spinor}) through the upper one and following
\cite{Popov-center,Popov}, we obtain an effective second order differential
equation in the form of the Schr{\"o}dinger equation
\be
\chi^{\prime\prime}(r)+k^2(r)\chi(r)=0, \quad\quad\quad
a(r)=\exp\left[\frac{1}{2}\int(\frac{1}{r}-\frac{\tilde{V}^{\prime}}{\epsilon+m-\tilde{V}})\,dr\right]\,\chi(r).
\label{Schrodinger-like:eq}
\ee
Here
\be
k^2(r)=2({\cal E}-U(r)),\quad {\cal
E}=\frac{\epsilon^2-m^2}{2},\quad \tilde{V}=\frac{V}{\hbar v_F}
\ee
and we
represent the effective potential as the sum of two terms $U=U_{1}+U_{2}$,
where $U_1$ is the effective potential for the Klein--Gordon equation and $U_2$
takes into account the spin dependent effects
\ba
U_{1}&=&\epsilon\tilde{V}-\frac{\tilde{V}^{2}}{2}+\frac{j(j-1)}{2r^{2}},
\label{U1-pot}\\
U_{2}&=&\frac{1}{4}\left[\frac{\tilde{V}^{\prime\prime}}{\epsilon+m-\tilde{V}}+\frac{3}{2}
\left(\frac{\tilde{V}^{\prime}}{\epsilon+m-\tilde{V}}\right)^{2}+\frac{2j\tilde{V}^{\prime}}
{r(\epsilon+m-\tilde{V})}\right]. \label{U2-pot}
\ea
Note that Eq.(\ref{Schrodinger-like:eq}) and the potentials (\ref{U1-pot}),
(\ref{U2-pot}) coincide with the corresponding equations in QED \cite{Popov}.
We plot the effective potential $U(r)$ for $Z \to Z_c$, $j=1/2$, and
$\epsilon=-m$ in Fig.\ref{coul1}, where the Coulomb barrier is clearly seen.
\begin{figure}[h]
\includegraphics[width=6.5cm]{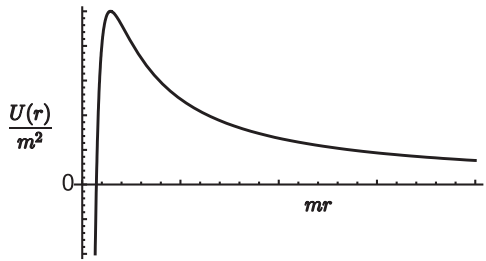}
\caption{Effective potential for the Coulomb center in the case $\epsilon=-m$
and $Z=Z_{c}$.}
\label{coul1}
\end{figure}

Up to now, we considered a one particle problem in an external field. In the
next section, we will consider electrons and holes in graphene which interact
by means of the Coulomb interaction and show that an instability develops in
the system when the coupling $\alpha$ exceeds some critical value $\alpha_{c}$.

\section{The excitonic tachyon instability}
\label{excitonic-instability}

\subsection{The Bethe--Salpeter equation}

The instability of the supercritical charge problem due to the emission of
positively charged holes discussed in the previous section indicates the
possibility of the excitonic instability in graphene in the case of a
supercritical coupling constant. In this section, we will study the
Bethe-Salpeter (BS) equation for an electron-hole state and show that it has a
tachyon in the spectrum in the supercritical regime. Before we do this, let us
discuss some similarities and differences of the supercritical Coulomb charge
problem with the famous Cooper problem in the theory of superconductivity.

Although the Cooper problem is formulated as a quantum mechanical problem for
two particles (electrons), it can be standardly reduced to a one
particle problem in an external potential. Therefore, the Coulomb center
problem is similar to the Cooper problem in this respect. However, there are
important differences between the two problems. The first one is connected with
the fact that the Dirac equation contains the lower continuum  with filled
negative energy states. Therefore, if a bound state energy enters the lower
continuum, we are essentially dealing with a many body problem. This explains
why there are resonant states with imaginary energy in the supercritical Coulomb
potential unlike the Cooper problem where there are only negative energy bound
states. The second important difference between these two problems is connected
with the critical value of coupling constant. It is zero for the Cooper problem
because the density of states in this problem is nonzero at the Fermi surface
that plays a crucial role in the bound states formation. On the other hand,
$\alpha_{c}=1/2$ for the Coulomb center problem in graphene where the density
of states is zero at the Dirac point.

The appearance of the Cooper bound state  in the theory of
superconductivity is directly related to the instability of the normal state of
metal. Indeed, according to \cite{Schrieffer}, the BS equation for
an electron-electron bound state in the normal state of metal has a
solution with imaginary energy, i.e. a tachyon. This means that normal state is unstable and a phase
transition to the superconducting state takes place. As we mentioned above,
resonant states in the supercritical Coulomb center problem suggest the
excitonic instability in graphene.

For the description of the dynamics in graphene, we will use the
same model as in Refs.\cite{Khveshchenko,Gra2002}  in which while quasiparticles are confined to
a two-dimensional plane,  the electromagnetic (Coulomb) interaction between
them is three-dimensional in nature. The low-energy quasiparticles excitations in
graphene are conveniently described in terms of a four-component Dirac spinor
$\Psi^{T}_{a}=(\psi_{KAa},\psi_{KBa},\psi_{K^{\prime}Ba},\psi_{K^{\prime}Aa})$
which combines the Bloch states with spin indices $a=1,2$
on the two different sublattices $(A, B)$ of the hexagonal graphene lattice and
with momenta near the two nonequivalent valley points $(K, K^{\prime})$ of the
two-dimensional Brillouin zone. In what follows we treat the spin index as a ``flavor''
index with $N_{f}$ components, $a=1,2,\dots N_{f}$, then $N_{f}=2$ corresponds
to graphene monolayer while $N_{f}=4$ is related to the case of two
decoupled graphene layers, interacting solely via the Coulomb
interaction.

The action describing graphene quasiparticles interacting
through the Coulomb potential has the form
\ba
S&=&\int dtd^{2}r\overline{\Psi}_{a}(t,{\mathbf{r}})\left(i\gamma^{0}\partial_{t}-iv_{F}
\mathbf{\gamma}\mathbf{\nabla}\right)\Psi_{a}(t,\mathbf{r})\nonumber\\
&-&\frac{1}{2}\int dt dt^{\prime}
d^{2}rd^{2}r^{\prime}\overline\Psi_{a}(t,\mathbf{r})\gamma^{0}\Psi_{a}(t,\mathbf{r})
U_{0}(t-t^{\prime},|\mathbf{r}-\mathbf{r}^{\prime}|)
\overline\Psi_{b}(t^{\prime},\mathbf{r}^{\prime})\gamma^{0}\Psi_{b}(t^{\prime},\mathbf{r}^{\prime}),
\label{action}
\ea
where $\overline\Psi=\Psi^{\dagger}\gamma^{0}$, and the $4\times4$ Dirac $\gamma$-matrices
$\gamma^{\mu}=\tau^3\bigotimes \, (\sigma^3,i\sigma^2,-i\sigma^1)$ furnish a reducible
representation of the Dirac algebra in $2+1$ dimensions. The Pauli matrices $\mathbf{\tau},
\mathbf{\sigma}$ act in the subspaces of the valleys ($K,K^{\prime}$) and sublattices ($A,B$), respectively.
The other two $\gamma$-matrices which we use are $\gamma^{3}=i\tau_{2}\otimes\sigma_{0},
\gamma^{5}=i\gamma^{0}\gamma^{1}\gamma^{2}\gamma^{3}
=\tau_{1}\otimes\sigma_{0}$ ($\sigma_{0}$ is the $2\times2$ unit matrix).

The bare Coulomb potential
$U_{0}(t,|\mathbf{r}|)$ takes the simple form:
\be
U_{0}(t,|\mathbf{r}|)=\frac{e^{2}\delta(t)}{\kappa}\int \frac{d^{2}k}{2\pi}\frac{e^{i\mathbf{k}\mathbf{r}}}
{|\mathbf{k}|}=\frac{e^{2}\delta(t)}{\kappa|\mathbf{r}|}.
\ee
However, the polarization effects considerably modify this bare Coulomb potential and the interaction
will be
\be
U(t,|\mathbf{r}|)=\frac{e^{2}}{\kappa}\int\frac{d\omega}{2\pi}\int \frac{d^{2}k}{2\pi}
\frac{\exp(-i\omega t +i\mathbf{k}\mathbf{r})}{|\mathbf{k}|+\Pi(\omega,\mathbf{k})},
\ee
where $\kappa$ is the dielectric constant due to a substrate and the polarization function $\Pi(\omega,\mathbf{k})$
is proportional (within the factor $2\pi/\kappa$) to the time component of the photon polarization function.
Correspondingly,  the Coulomb propagator has the form
\be
D(\omega,|\mathbf{q}|)=\frac{1}{|\mathbf{q}|+\Pi(\omega,|\mathbf{q}|)}.
\label{Coulombpropagator}
\ee
The one-loop polarization function is \cite{Gonzalez}
\be
\Pi(\omega,\mathbf{k})=\frac{\pi e^{2}N_{f}}{4\kappa}\frac{\mathbf{k}^{2}}
{\sqrt{\hbar^{2}v_{F}^{2}\mathbf{k}^{2}-\omega^{2}}},
\label{1loop-polarization}
\ee
and in an instantaneous approximation it is
\be
\Pi(\omega=0,\mathbf{k})=\frac{\pi e^{2}N_{f}}{4\kappa\hbar v_{F}}|\mathbf{k}|.
\ee
In general, the static polarization operator must have the form
$\Pi(0,|\mathbf{q}|)=|\mathbf{q}|F(\alpha,N_{f})$ due to dimensional reasons,
however its exact form is not known and in the present paper we will use the one-loop
approximation.

The continuum effective theory described by the action (\ref{action}) possesses the
$U(2N_{f})$ symmetry. However, as was pointed out in Ref. \cite{Alicea} (see
also Refs. \cite{Herbut,Aleiner}), it is not exact for the Lagrangian on the
graphene lattice. In fact, there are small on-site repulsion interaction terms
which break the $U(2N_{f})$ symmetry.

\begin{figure}[t]
\includegraphics[width=8.5cm]{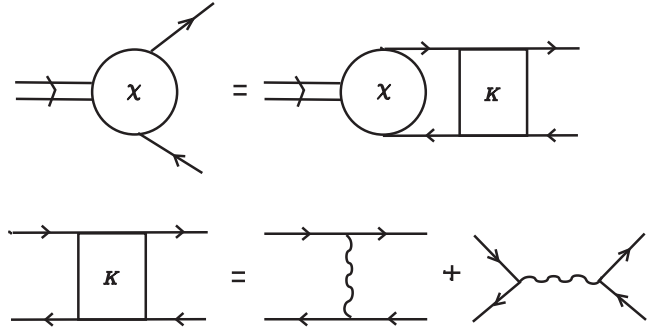}
\caption{The BS equation for a bound electron-hole state $\chi$. The kernel
$K$ contains two diagrams: exchange and annihilation ones. The wave line
corresponds to the Coulomb propagator.}
\label{BS:eq}
\end{figure}

In order to analyze excitonic instability, we consider the
Bethe--Salpeter  equation (see, for example, Ref.\cite{Miranskybook})
for an electron-hole bound state which is represented
in Fig.\ref{BS:eq}.
The kernel $K$ of the BS equation in the simplest approximation contains two
diagrams: the one is due to exchange Coulomb forces and another one is
the annihilation diagram. The annihilation
diagram does not contribute for the BS wave function considered
below. Thus the BS equation takes the following form:
\be
\left[S^{-1}(q+\frac{1}{2}P)\chi(q,P)S^{-1}(q-\frac{1}{2}P)\right]_{\alpha\beta}
=\frac{i\alpha}{(2\pi)^{2}}\int d^3k\,D(|\mathbf{q}-\mathbf{k}|)
\left[\gamma^0\chi(k,P)\gamma^0\right]_{\alpha\beta},
\label{BS1}
\ee
where $k=(k_0,\mathbf{k})$, $\alpha,\beta$ are spinor indices,
$\chi(q,P)$ is the BS amplitude in momentum space
\be
\chi_{\alpha\beta}(q,P)=\int
d^{3}x\,e^{iqx}\langle0|T\Psi_{\alpha}\left(\frac{x}{2}\right)
\overline{\Psi}_{\beta}\left(-\frac{x}{2}\right)|P\rangle,
\ee
$q=(q_0,\mathbf{q})$, $P=(P_0,\mathbf{P})$, $\mathbf{q}$ and $\mathbf{P}$ are relative and
total momenta, respectively,  and
$$
S(p)=\frac{\gamma^{0}p_{0}-\mathbf{\gamma}\mathbf{p}+\Delta}{p_0^2-\bp^2-\Delta^2 +i0}
$$
is the quasiparticle propagator with a gap $\Delta$ (the gap $\Delta$ is zero in
non-interacting graphene, however, may be generated due to the strong Coulomb
interaction). In what follows we put $\hbar=v_{F}=1$.

Taking into account the static vacuum polarization by massless fermions, i.e.,
with $\Pi(\omega=0,\mathbf{k})$,
corresponds to the replacement of the coupling constant $\alpha$ in Eq.(\ref{BS1}) by
$$
\alpha \rightarrow \frac{\alpha}{1+\pi \alpha N_{f}/4}\equiv 2\lambda.
$$
Further, introducing the function
$$
\widehat{\chi}(q,p) = S^{-1}(q+\frac{1}{2}P)\chi(q,P)S^{-1}(q-\frac{1}{2}P)
$$
with $p=P/2$, the BS equation can be equivalently rewritten as
follows:
\be
\widehat{\chi}(q,p)=\frac{2i\lambda}{(2\pi)^{2}}\int\frac{d^{3}k}{|\mathbf{q}-\mathbf{k}|}
\gamma^0 S(k+p)\widehat{\chi}(k,p)S(k-p)\gamma^0.
\label{BS2}
\ee
In general, $\widehat{\chi}$ can be expanded in 16 independent matrix
structures. In view of the experience in QED \cite{review}, we
expect a gap generation in graphene in the supercritical regime. Then the
spin-valley $U(4)$ symmetry  will be broken (see, e.g. \cite{Khveshchenko,Gra2002}) that leads
to the appearance of massless Nambu--Goldstone bosons in the spectrum.
Similarly to QED \cite{review}, these Nambu--Goldstone bosons are
transformed into tachyons if considered on the wrong vacuum state without a gap
generation. In the present paper, we will consider only matrix structures of
$\widehat{\chi}$ connected with the $\gamma^5$ matrix
\be
\widehat{\chi}(q,p) =
\chi_5(q,p)\gamma^5 +\chi_{05}(q,p)
q^i\gamma^i\gamma^0\,\gamma^5,
\label{general-expansion}
\ee
where $\chi_5(q,p)$ and $\chi_{05}(q,p)$ are scalar coefficient functions. We will see in the
next section that it is enough to consider only $\chi_5$ in order to describe a
Nambu--Goldstone excitation in the massive state. However, we retain the
function $\chi_{05}$ because it is necessary in the study of tachyon.
There are also tachyons in other channels which
describe different ways of breaking the $U(2N_{f})$ symmetry, for example, one can use
matrices $I,\gamma^{3}, \gamma^{3}\gamma^{5}$ instead of the matrix $\gamma^{5}$
in Eq.(\ref{general-expansion}). To study instability it is enough to find at least one channel
with tachyons. The real pattern of
a symmetry breaking is defined by solving gap equations for various kinds of
order parameters and determining which of them corresponds to the global minimum of
the system energy.
For simplicity we consider only the channel described by the wave function
(\ref{general-expansion}) which can be treated analytically.

\subsection{Tachyon states}

Let us first show that, for $\lambda > \lambda_{c}$, there is a tachyon in the spectrum of
the Bethe--Salpeter equation in the massless theory $\Delta=0$ and determine the
critical value $\lambda_{c}$. For the study of tachyon, we can set $\bp=0$,
however, should keep nonzero $p_0$.  One can check that ansatz
(\ref{general-expansion}) is consistent for Eq.(\ref{BS2}) and leads to a coupled
system of equations for functions $\chi_5(q,p_{0}),\chi_{05}(q,p_{0})$ (in what follows
we omit $p_{0}$ for brevity in the arguments of the functions $\chi_5(q,p_{0}),\chi_{05}(q,p_{0})$).
 Since Eq.(\ref{BS2}) implies
that $\widehat{\chi}(q,p)$ does not depend on $q_0$, we can
integrate then over $k_0$ by using the integrals
$$
i\int\limits_{-\infty}^{\infty}\,\frac{dk_0}{\pi}\,
\frac{c_1+c_2\,k_0+c_3k_0^2}{\left((k_0-p_0)^2-\mathbf{k}^2+i\delta\right)
\left((k_0+p_0)^2-\mathbf{k}^2+i\delta\right)}\,\,
 =\,\, \frac{c_1+c_3(p^2_0-\mathbf{k}^2)}{2|\mathbf{k}|\left(p_0^2-\mathbf{k}^2\right)}\,,
$$
where $\delta \to +0$. We obtain the following system of integral equations:
\be
\chi_5(\mathbf{q}) = \lambda\int
\frac{d^2k}{2\pi}\frac{\bk^2\left(\chi_5(\mathbf{k})+p_0\chi_{05}(\mathbf{k})\right)}
{|\bq-\bk||\bk|\left(\bk^2-p_0^2\right)}, \quad\quad\quad\quad \chi_{05}(\mathbf{q}) =
{\lambda}\int
\frac{d^2k}{2\pi}\frac{{\bq\bk}\left(\bk^2\chi_{05}(\mathbf{k})+
p_0\chi_5(\mathbf{k})\right)}{\bq^2|\bq-\bk||\bk|\left(\bk^2-p_0^2\right)}.
\label{eq-tachyon}
\ee
We assume that $\chi_5(\mathbf{q})$ and $\chi_{05}(\mathbf{q})$ depend only on
$q=|\mathbf{q}|$, then we can integrate over the angle using
\ba
\int\limits_{0}^{2\pi} \frac{d\phi}{\sqrt{q^2+k^2-2qk\cos\phi}} &=&
\frac{4}{q+k}K\left(\frac{2\sqrt{qk}}{q+k}\right)=4\left[\frac{\theta(q-k)}{q}
K\left(\frac{k}{q}\right)+\frac{\theta(k-q)}{k}
K\left(\frac{q}{k}\right)\right],
\label{identity1}\\
\int\limits_{0}^{2\pi} \frac{d\phi\cos\phi}{\sqrt{q^2+k^2-2qk\cos\phi}} &=&
\frac{2(q^2+k^2)}{qk(q+k)}\left(K\left(\frac{2\sqrt{qk}}{q+k}\right)
-\frac{(q+k)^2}{q^2+k^2}E\left(\frac{2\sqrt{qk}}{q+k}\right)\right)\nonumber\\
&=&\frac{4}{qk}\left[q\theta(q-k)\left(K\left(\frac{k}{q}\right)-E\left(\frac{k}{q}
\right)\right)+k\theta(k-q)\left(K\left(\frac{q}{k}\right)-E\left(\frac{q}{k}
\right)\right)\right],
\label{identity2}
\ea
where $K(x)$ and $E(x)$ are complete elliptic integrals of the first and second kind, respectively,
$\theta(x)$ is the Heaviside step function, and for  the last equalities in
Eqs.(\ref{identity1}), (\ref{identity2}) we used the formulae $8.126.3, 8.126.4$
in the book \cite{GR}. Further, we approximate the elliptic integrals by their
asymptotics
\be
K(x)\simeq\frac{\pi}{2}\left(1+O(x^{2})\right),\quad
E(x)\simeq\frac{\pi}{2} \left(1+O(x^{2})\right),\quad x\ll1.
\label{approximation}
\ee
This approximation allows one to obtain analytical results for the BS equation.
The logarithmic singularity present in elliptic integrals in Eqs.(\ref{identity1}), (\ref{identity2}) at $q=k$
does not influence qualitatively the solution obtained though it is important to take it
into account to get correct value of the critical coupling (see, the derivation of  Eq.(\ref{exact-crit-coupling})
below).
Thus we find
\ba
\chi_5(q) &=&\lambda\int\limits_{0}^q
\frac{k^2dk}{q(k^2-p_0^2)}\left(\chi_5(k)+p_0\chi_{05}(k)\right)+
\lambda\int\limits^{\Lambda}_q
\frac{kdk}{k^2-p_0^2}\left(\chi_5(k)+p_0\chi_{05}(k)\right),\\
\chi_{05}(q) &=& \frac{\lambda}{2}\int\limits_{0}^q
\frac{k^2dk}{q^3(k^2-p_0^2)}\left(k^2\chi_{05}(k)+ p_0\chi_5(k)\right)+
  \frac{\lambda}{2}\int\limits^{\Lambda}_q \frac{dk}{k(k^2-p_0^2)}\left(k^2\chi_{05}(k)+
  p_0\chi_5(k)\right).
\ea
Here we also introduced a finite ultraviolet cutoff $\Lambda$ which could
be taken to be of order $\pi/a$, where $a$ is a characteristic lattice size,
$a=2.46 {\AA}$ for graphene. An alternative, equally good, choice
of $\Lambda$ is related to the energy band, $\Lambda=t/v_{F}$,
where $t=2.4$ eV in  graphene.

These equations are equivalent to the system of differential equations
\be
\chi_5''+\frac{2}{q}\chi_5'+\lambda\frac{\chi_5+p_0\chi_{05}}{q^2-p_0^2}=0,
\quad\quad\quad\quad
\chi_{05}''+\frac{4}{q}\chi_{05}'+\frac{3\lambda}{2}\frac{q^2\chi_{05}+ p_0
\chi_5}{q^2(q^2-p_0^2)} = 0
\label{coupled-system}
\ee
with the following boundary conditions:
\be
q^{2}\chi'_5\Big|_{q=0}=0,\quad
[q\chi_5(q)]^{\prime}\Big|_{q=\Lambda}=0, \quad
q^{4}\chi_{05}'\Big|_{q=0}=0,\quad
[q^3\chi_{05}(q)]^{\prime}\Big|_{q=\Lambda}=0.
\label{boundary-conditions}
\ee
The system of differential equations (\ref{coupled-system}) can be reduced
to one equation of the fourth
order whose solutions are given in terms of generalized hypergeometric
functions $_{4}F_{3}(q^{2}/p^{2}_{0})$ and the Meijer functions with the corresponding boundary
conditions (for this analysis see appendix \ref{Pochhammer:eq}).
However, since we seek for the solution with  ${p}_0
\rightarrow 0$, it is simpler to analyze straightforwardly the system
(\ref{coupled-system}) itself, in this regime the system decouples
\be
\chi_5''+\frac{2}{q}\chi_5'+\lambda\frac{\chi_5}{q^2-{p}_0^2}=0,
\quad\quad\quad\quad
\chi_{05}''+\frac{4}{q}\chi_{05}'+\frac{3\lambda}{2}\frac{\chi_{05}}{q^2-{p}_0^2}
= 0, \quad\quad\quad\quad
\label{decoupled-system}
\ee
 where we keep ${p}_0$ in the denominators because it regularizes singularities
 for $q \to 0$.

Obviously, Eqs.(\ref{decoupled-system}) are differential equations for the
hypergeometric function $F(a,b;c;z)$ \cite{GR}. The solutions that satisfy the
infrared boundary conditions are
\be
\chi_5=C_1F\left(\frac{1+\gamma}{4},\frac{1-\gamma}{4};\frac{3}{2};\frac{q^2}{p_0^2}\right),\,\,\,\,\,
\chi_{05}=C_2F\left(\frac{3(1+\tilde{\gamma})}{4},\frac{3(1-\tilde{\gamma})}{4};
\frac{5}{2};\frac{q^2}{p_0^2}\right),
\ee
where $\gamma=\sqrt{1-4\lambda}$ and $\tilde{\gamma}=\sqrt{1-2\lambda/3}$.
Using the asymptotic of the
hypergeometric functions, one may easily check that the ultraviolet boundary
conditions for the function $\chi_5$ can be satisfied only for $\lambda>1/4$,
therefore, $1/4$ is the critical coupling for the approximation that we
use. (Note that if we neglect the vacuum polarization contribution, then
$\lambda=\alpha/2$ and the critical value $1/4$ coincides with the critical
coupling $Z_c\alpha=1/2$ obtained in Sec.II for the Coulomb center problem.)
The UV boundary condition for the function $\chi_{05}$ can be satisfied for the
values of $\lambda>3/2$ but not for $\lambda<3/2$. Therefore, for
$1/4<\lambda<3/2$ we take a trivial solution $\chi_{05}=0$ and we are left only
with the equation for the function $\chi_5$. Knowing the function $\chi_5$ we
then solve an inhomogeneous equation (\ref{coupled-system}) for $\chi_{05}$, in
this way we find that the function $\chi_{05}\sim p_{0}$. The critical
value $\lambda_c=1/4$ coincides with the critical coupling constant found in
\cite{Gra2002}, where the same approximation for the kernel was made. In the
supercritical regime $\gamma = i\omega$, $\omega=\sqrt{4\lambda-1}$ and the
function $\chi_{5}(q)$ behaves asymptotically as
\be
\chi_{5}(q)\sim
q^{-1/2}\cos\left(\sqrt{\lambda-1/4}\ln q +const\right).
\ee
 Such oscillatory behavior is typical for the phenomenon known
in quantum mechanics as the collapse (``fall into the center'') phenomenon: in
this case the energy of a system is unbounded from below and there is no ground
state. Nodes of the wave function of the bound state signify the existence of
the tachyon states with imaginary energy $p_{0}$, ${\rm Im}p_{0}^{2}<0$. Indeed, the UV
boundary condition for $\chi_5$ leads to the equation
\be
\frac{(1+i\omega)\Gamma\left(1+\frac{i\omega}{2}\right)\Gamma\left(\frac{1-i\omega}{4}\right)
\Gamma\left(\frac{5-i\omega}{4}\right)}
{(1-i\omega)\Gamma\left(1-\frac{i\omega}{2}\right)
\Gamma\left(\frac{1+i\omega}{4}\right)\Gamma\left(\frac{5+i\omega}{4}\right)}
\left(-\frac{\Lambda^2}{p_0^2}\right)^{i\frac{\omega}{2}}=1.
\ee
Then we find the following tachyon solution:
\be
p^2_0 = -\Lambda^2\exp\left(-\frac{4\pi n}{\omega}+\delta(\omega)\right),\quad \delta(\omega)=
\frac{4}{\omega}\left[\arctan\omega+{\rm Arg}\left(\Gamma\left(1+\frac{i\omega}{2}\right)
\Gamma\left(\frac{1-i\omega}{4}\right)
\Gamma\left(\frac{5-i\omega}{4}\right)\right)\right].
\ee
If $\lambda$ tends to $1/4$ from the above, i.e. $\omega \rightarrow 0$,
\be
p^2_0 = -\Lambda^2\exp\left(-\frac{4\pi
n}{\omega}+\delta(0)\right),\quad \delta(0)=
4+2\Psi(1)-\Psi(1/4)-\Psi(5/4)\approx7.3,\quad n=1,2,\dots.
\label{tachyon}
\ee
Thus, we see that the strongest instability, i.e., the smallest negative value of $p_0^2$ is
given by the solution for the function $\chi_5$ with $n=1$. The tachyon  states
play here the role of the quasistationary states in the problem of
supercritical Coulomb center resulting in the vacuum instability. In fact, the
tachyon instability can be viewed as the field theory analogue of the ``fall
into the center'' phenomenon and the critical coupling $\alpha_{c}$ is an
analogue of the critical coupling $Z_{c}\alpha$ in the problem of the Coulomb
center.

The  tachyon energy Eq.(\ref{tachyon}) has a characteristic essential singularity
of the kind $1/\sqrt{\lambda-\lambda_{c}}$ in the exponent. It can be argued that this
behavior reflects a scale invariance in the problem under consideration and keeps its form
for any approximation which does not introduce new scale parameter except the cutoff
\cite{Holdom}.

There are two possibilities for the system with the supercritical charge to
become stable: to shield spontaneously the charge or to generate spontaneously
the fermion gap. The first possibility is realized in the problem of the
supercritical Coulomb center which is due to the formulation of the problem as
the one-particle one. The second possibility - dynamical generation of the
fermion gap - is realized for quasiparticles in graphene interacting through
supercritical Coulomb interaction. The situation here is completely analogous
to the strongly coupled QED \cite{review,Fomin,Miranskybook} where it is shown
that the vacuum stabilization by generating dynamical fermion gap is a rather
universal phenomenon.

The critical value $\lambda_{c}$ determines the critical coupling $\alpha_{c}$
as a function of the fermion number $N_{f}$, \be
\alpha_{c}=\frac{4\lambda_{c}}{2-\pi N_{f}\lambda_{c}},
\label{critline-Coulomb} \ee (compare with Eq.(28) in \cite{Gra2002}). The
critical value $\lambda_{c}=1/4$ in the approximation (\ref{approximation})
used for kernels. The more precise value of $\lambda_{c}$ can be found if one
notes that $\lambda_{c}$ corresponds to the limit $p_{0}=0$. Taking this limit
in the system (\ref{coupled-system}), we get \ba
\chi_{5}(q)&=&\frac{2\lambda}{\pi}\int\limits_{0}^{\infty}d k
\chi_{5}(k)\left[\frac{\theta(q-k)}{q}
K\left(\frac{k}{q}\right)+\frac{\theta(k-q)}{k}
K\left(\frac{q}{k}\right)\right],\\
\chi_{05}(q)&=&\frac{2\lambda}{\pi q}\int\limits_{0}^{\infty}d k \chi_{05}(k)\left[\theta(q-k)
\left(K\left(\frac{k}{q}\right)-E\left(\frac{k}{q}\right)\right)+\frac{k\theta(k-q)}{q}
\left(K\left(\frac{q}{k}\right)-E\left(\frac{q}{k}\right)\right)\right].
\ea
Note that the ultraviolet cutoff, $\Lambda$, has been taken to infinity, which is appropriate
at the critical point. These equations are scale invariant and  are solved by
$\chi_{5}(q)=q^{-\gamma}, \chi_{05}(q)=q^{-\rho}$
on the condition that the exponents $\gamma,\rho$ satisfy the transcendental equations
\ba
1&=&\frac{2\lambda}{\pi}\int\limits_{0}^{1}dx\left[x^{-\gamma}+x^{\gamma-1}\right]K(x),\quad
0<\gamma<1,
\label{trans:eq1}\\
1&=&\frac{2\lambda}{\pi}\int\limits_{0}^{1}dx\left[x^{-\rho}+x^{\rho-3}\right](K(x)-E(x)),
\quad 0<\rho<3. \label{trans:eq2} \ea These equations define roots
$\gamma,\rho$ for any value of the coupling $\lambda$. An instability is
signalized by oscillatory behavior of the functions $\chi_{5}(q),\chi_{05}(q)$.
For the function $\chi_{5}(q)$ this occurs when two of the roots of
Eq.(\ref{trans:eq1}) in the interval $(0,1)$ coalesce and then become complex conjugate. We
find that this happens when $\gamma=1/2$, for this value the integral in
Eq.(\ref{trans:eq1}) is exactly evaluated (see, the book
\cite{Prudnikov}) and we obtain the critical value
\be
\lambda_{c}=\frac{4\pi^{2}}{\Gamma^{4}(1/4)}\approx0.23.
\label{exact-crit-coupling}
\ee
The second equation
(\ref{trans:eq2}) gives higher critical value $\lambda_{c}=0.91$ therefore the
instability is determined by the value $\lambda_{c}=0.23$. The critical value
$N_{crit}\approx2.8$ corresponds to $\alpha=\infty$ in
Eq.(\ref{critline-Coulomb}). Since for graphene the number of ''flavors``
$N_{f}=2$, the critical coupling is estimated to be $\alpha_{c}\approx 1.62$
in the considered approximation \footnote{Equivalently,
the critical coupling can be determined from a gap equation. In Ref. \cite{Gra2002}
some approximation for the kernel like in Eq.(\ref{approximation}) was used
for the gap equation (\ref{eq:Delta})
that gave an overestimated value $\alpha_{c}$  ($\alpha_{c}=2.33$ there).
The numerical analysis of Eq.(\ref{eq:Delta}) in the second paper in \cite{Khveshchenko}
yielded $\alpha_{c}=1.1$.}. Because the coupling constant in freely standing
graphene $\alpha\approx2.19$ ($\kappa\approx1$) the system
is in the unstable phase. On the other hand, for graphene on a SiO$_{2}$ substrate the
dielectric constant $\kappa\approx2.8$, therefore, $\alpha\approx0.78$, i.e.,
the system is in the stable phase.

Finally, since the $U(2N_{f})$ symmetry is spontaneously broken, there must exist
Nambu--Goldstone excitations in the stable phase where a quasiparticle gap
arises. Let us show that the BS equation (\ref{BS2}) indeed admits
such solutions. To see this, according to \cite{review}, we set $p_0=\bp=0$.
Then, Eq.(\ref{BS2}) has a solution of the form
$\chi(q,0)=\chi_5(q,0)\gamma_{5}$ for which we obtain the equation
\be
\label{CHI5} \chi_5(q,0) = \frac{\lambda}{2\pi}\int
\frac{d^2k}{|\bq-\bk|}\frac{\chi_5(k,0)}{\sqrt{\bk^2+\Delta^2(k)}},
\ee
or, after integrating over the angle,
\be \chi_5(q,0) =
\lambda\int\limits_{0}^{\Lambda} \frac{d k\,
k\chi_5(k,0)}{\sqrt{k^2+\Delta^2(k)}} {\cal K}(q,k), \label{eq:chi}
\ee
with the kernel
\be
{\cal K}(q,k)=\frac{\theta(q-k)}{q}K\left(\frac{k}{q}\right)+\frac{\theta(k-q)}{k}
K\left(\frac{q}{k}\right).
\ee
On the other hand, the equation for a gap
function obtained in Ref. \cite{Gra2002} has the form
\be
\Delta(q)=\lambda\int\limits_{0}^{\Lambda} \frac{d k\,
k\Delta(k)}{\sqrt{k^2+\Delta^2(k)}} {\cal K}(q,k).
\label{eq:Delta}
\ee
One can see that the equation (\ref{eq:chi}) has the solution $\chi_5(q,0)=C\Delta(q)$
where the gap function $\Delta(q)$ satisfies the equation (\ref{eq:Delta}) and $C$
is a constant. Thus the wave function $\chi_5(q,0)$ describes a gapless
Nambu--Goldstone excitation. Solving the BS equation at nonzero
$p_{0},\mathbf{p}$ one can obtain a dispersion law $p_{0}\sim|\mathbf{p}|$ for
a Nambu--Goldstone excitation.

\section{Conclusions}
In this paper we studied instabilities in graphene which arise at strong
Coulomb coupling. For the supercritical Coulomb center problem, it
was known before that the ``fall into the center" instability arises if
$Z\alpha$ exceeds the critical value $1/2$ leading to the appearance of
quasistationary levels with complex energies.
The energy of quasistationary states  in the case of gapless quasiparticles has a characteristic
essential-singularity type dependence on the coupling constant reflecting the
scale invariance of the Coulomb potential. We showed that a quasiparticle
gap stabilizes the system decreasing the imaginary part $|{\rm Im}E|$
of quasistationary states, thus increasing their lifetime.

Considering the
many-body problem of strongly interacting gapless quasiparticles in graphene,
we showed that the Bethe-Salpeter equation for an electron-hole bound state
contains a tachyon in its spectrum in the supercritical regime
$\alpha>\alpha_{c}$ and found the critical constant $\alpha_{c}=1.62$ in the
static random phase approximation. The tachyon states play the role of
quasistationary states in the problem of the supercritical Coulomb center and
lead to the rearrangement of the ground state and the formation of exciton
condensate. Thus, there is a close relation between the two instabilities,  in
fact, the tachyon instability can be viewed as the field theory analogue of the
``fall into the center'' phenomenon and the critical coupling $\alpha_{c}$ is
an analogue of the critical coupling $Z_{c}\alpha$ in the problem of the
Coulomb center. The physics of two instabilities is related to strong Coulomb
interaction.

The calculated critical value $\alpha_{c}=1.62$ should be compared with the
value $\alpha_{c}=1.08$ found in Monte Carlo simulations \cite{Drut} for the
rearrangement of the ground state of graphene and appearance of a gap. The obtained value of
$\alpha_{c}$ is rather large that indicates that the ladder approximation is
not quantitatively good enough for the problem of excitonic instability and gap
generation in freely standing graphene. Certainly, both higher order
corrections and improving the instantaneous approximation can vary the value of
critical coupling. It is essential however that a ground state rearrangement at
strong coupling is connected with the ``fall into the supercritical Coulomb
center'' phenomenon. Therefore, such an rearrangement in graphene
with large Coulomb interaction seems to be very plausible for strong enough
coupling even if one goes beyond the ladder approximation. Finally, the
physical picture of instabilities in graphene is quite similar to that
elaborated earlier in strongly coupled QED \cite{Fomin,review,Miranskybook}
(see, also, \cite{Bardeen,Kogut}). In QED, the ladder approximation is not
reliable quantitatively also because the critical coupling constant for chiral
symmetry breaking is of order one. However, the main results of the
ladder approximation survive when all diagrams with photons exchanges are
included (the so-called quenched approximation without fermion loops)
\cite{Holdom}. Further, the existence of the critical point is exactly proved
in the lattice version of QED \cite{Seiler}. We note also that in the
presence of an external magnetic field the value of the critical coupling
reduces to zero (magnetic catalysis phenomenon \cite{magcatalysis}) so that the gap
generation takes place already in the weak coupling regime.

\begin{acknowledgments}
We are grateful to V.A. Miransky and S. Sharapov for useful discussions. The
work of E.V.G and V.P.G. was supported partially by
Ukrainian State Foundation for Fundamental Research under Grant
No. F28.2/083, by the grant 10/07-N
``Nanostructure systems, nanomaterials, nanotechnologies'',
and by the Program of Fundamental Research of the Physics
and Astronomy Division of the National Academy of Sciences
of Ukraine.

\end{acknowledgments}

\appendix
\section{Discrete spectrum for a regularized Coulomb potential}
\label{discretespectrum}

\begin{figure}[t]
\includegraphics[width=6.5cm]{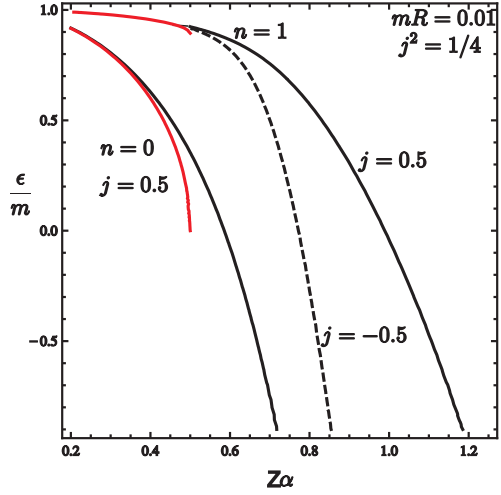}
\caption{The lowest energy levels as functions of $Z\alpha$. Red lines
correspond to the pure Coulomb potential (they exist for $Z\alpha < 1/2$);
black solid lines are numerical solutions for $j=1/2,\, mR=0.01$; black dashed
line are numerical solutions for $j=-1/2,\, mR=0.01$. } \label{cr1}
\end{figure}

The discrete spectrum of Eq.(\ref{1}) exists for $|\e| < m$. In this case it is convenient to
define
\be
\rho =2 u r , \,\,\,\,\, u
=\sqrt{m^2-\e^2},\,\,\,\,\,\,\,\, a =\frac{\sqrt{m+\e}}{2}(g-f),\,\,\,\,\,\,b
=\frac{\sqrt{m-\e}}{2}(g+f) \label{ab}
\ee
and rewrite Eqs.($\ref{1}$) as follows:
\ba
\rho g' + g\left(\frac{\rho}{2}-\frac{1}{2}-Z\A\frac{\e}{u}\right)+
f\left(j+Z\A\frac{m}{u}\right) =0,\nonumber\\
\rho f' -
f\left(\frac{\rho}{2}+\frac{1}{2}-Z\A\frac{\e}{u}\right)+g\left(j-Z\A\frac{m}{u}\right)
=0. \label{system}
\ea
Substituting $f$ from the first equation into the second
one, we obtain the equation for the $g$ component
\be
\frac{d^2g}{d\rho^2} +
\left(-\frac{1}{4}+\frac{\frac{1}{2}+Z\A\frac{\e}{u}}{\rho}+\frac{\frac{1}{4}-j^2
+Z^2\alpha^2}{\rho^2}\right)g=0, \ee which is the well-known Whittaker equation
\cite{GR}. Its general solution is \be g =C_1 W_{\mu,\nu}[\rho] + C_2
M_{\mu,\nu}[\rho],\quad \mu=\frac{1}{2}+\frac{Z\alpha\e}{u},
\ee
where $\nu =\sqrt{j^2-Z^2\A^2}$. Taking into account the asymptotic of the Whittaker
functions $W_{k,\nu}(z),M_{k,\nu}(z)$ at infinity,
\ba
W_{\mu,\nu}(\rho)&\simeq& e^{-u r}(2u r)^{\mu},\quad r\to\infty,\\
M_{\mu,\nu}(\rho)&\simeq&\frac{\Gamma(1+\nu)}{\Gamma(\frac{1}{2}-\mu+\nu)}\,
e^{u r}(2u r)^{-\mu},\quad r\to\infty,
\ea
we find that the regularity condition at infinity requires $C_2 =0$. Then the first equation in
(\ref{system}) gives the following solution for the $f$ component in the region
II ($r > R$):
\be
f_{II}= C_1 \left(j-Z\A\frac{ m}{u}\right)W_{-\frac{1}{2}+Z\alpha\frac{\e}{u},\nu}[\rho].
\label{boundstate-wf}
\ee
Solutions in the region I ($r<R$) can be easily obtained from Eqs.($\ref{1}$)
\ba
b_I = A_1\,rJ_{|j+1/2|}\left(r\sqrt{\left(\e+\frac{Z\alpha}{R}\right)^2-m^2}\right),\\
a_I = A_1\,\mbox{sgn}(j)\,
\sqrt{\frac{\e+Z\alpha/R+m}{\e+Z\alpha/R-m}}\,\,rJ_{|j-1/2|}\left(r\sqrt{\left(\e+
\frac{Z\alpha}{R}\right)^2-m^2}\right),
\label{solutions-region-I}
\ea
where $A_1$ is a constant and we took into account the infrared boundary condition
which selects only regular solution for $b_I$ and $a_I$. Energy levels are
determined through the continuity condition of the wave function at $r=R$,
\be
\frac{b_I}{a_I}|_{r=R}=\frac{b_{II}}{a_{II}}|_{r=R},
\ee
that gives the equation
\be
\frac{W_{\frac{1}{2}+\frac{Z\alpha\epsilon}{u},\nu}(\rho)}{\left(j-\frac{Z\alpha
m}{u}\right)
W_{-\frac{1}{2}+\frac{Z\alpha\epsilon}{u},\nu}(\rho)}\Big|_{r=R} =
\frac{k+1}{k-1},\,\,\,\, k =
\mbox{sgn}(j)\,\frac{m+\e}{u}\sqrt{\frac{\e+Z\alpha/R-m}{\e+Z\alpha/R+m}}
\frac{J_{|j+1/2|}(\widetilde{\rho})}{J_{|j-1/2|}(\widetilde{\rho})},\,
\widetilde{\rho}=\sqrt{(Z\A+\e R)^2-m^2R^2}.
\label{matching}
\ee
We analyze this equation in the limit $R \to 0$ where we can use the asymptotical behavior
of the Whittaker function at $\rho\rightarrow0$,
\be
W_{\mu,\nu}(\rho)\simeq\frac{\Gamma(2\nu)}{\Gamma(\frac{1}{2}-\mu+\nu)}\rho^{\frac{1}{2}-\nu}
+\frac{\Gamma(-2\nu)}{\Gamma(\frac{1}{2}-\mu-\nu)}\rho^{\frac{1}{2}+\nu}.
\ee
In the limit $R \to 0$ Eq.(\ref{matching}) reduces to the
following one,
\ba \frac{\Gamma(-2\nu)}{\Gamma(2\nu)}\frac{\Gamma\left(1+\nu -
Z\A\frac{\e}{u}\right)}{\Gamma\left(1-\nu -
Z\A\frac{\e}{u}\right)}(2u R)^{2\nu}=
-\frac{j+\nu-\frac{Z\alpha(m+\epsilon)}{u}+k_{0}\left(j-\nu-\frac{Z\alpha(m-\epsilon)}{u}\right)}
{j-\nu-\frac{Z\alpha(m+\epsilon)}{u}+k_{0}\left(j+\nu-\frac{Z\alpha(m-\epsilon)}{u}\right)}+O(R),
\label{eq:Rtozero}
\ea
where
\be
k_{0}=\mbox{sgn}(j)\,\frac{m+\e}{u}\frac{J_{|j+1/2|}(Z\alpha)}{J_{|j-1/2|}(Z\alpha)}
\equiv \frac{m+\e}{u}\,\sigma(Z\alpha,j). \ee Using the relationships \be
\frac{j+\nu-\frac{Z\alpha(m-\epsilon)}{u}}{j-\nu-\frac{Z\alpha(m+\epsilon)}{u}}
=-\frac{Z\alpha}{j-\nu}\frac{u}{m+\epsilon},\quad
\frac{j-\nu-\frac{Z\alpha(m-\epsilon)}{u}}{j+\nu-\frac{Z\alpha(m+\epsilon)}{u}}
=-\frac{Z\alpha}{j+\nu}\frac{u}{m+\epsilon},
\ee
Eq.(\ref{eq:Rtozero}) can be rewritten in more convenient form
\be
\frac{\Gamma(-2\nu)}{\Gamma(2\nu)}\frac{\Gamma\left(1+\nu -
Z\A\frac{\e}{u}\right)}{\Gamma\left(1-\nu -
Z\A\frac{\e}{u}\right)}(2u R)^{2\nu}=
-\frac{j-\nu-\frac{Z\alpha(m-\epsilon)}{u}}{j+\nu-\frac{Z\alpha(m-\epsilon)}{u}}
\frac{j+\nu-Z\alpha\sigma(Z\alpha,j)}{j-\nu-Z\alpha\sigma(Z\alpha,j)}.
\label{eq:Rtozero2nd}
\ee
In the limit $R\to0$ the energy levels are determined
by the poles of the gamma function $\Gamma\left(1+\nu -
Z\A\frac{\e}{u}\right)$ and by a zero of the right hand side of
Eq.(\ref{eq:Rtozero2nd}), this leads to the familiar result (analogue of the
Balmer formula in QED) \cite{Khalilov} (rederived also in
\cite{Novikov}),
\be \e_{n,j}=
m\left[1+\frac{Z^2\A^2}{(\nu+n)^2}\right]^{-1/2},\quad \left\{\begin{array}{c}n=0,1,2,3,...,\, j>0,\\
\hspace{-2.2mm}n=1,2,3,...,\, \hspace{3mm}j<0.\end{array}\right.
\label{energy-levels}
\ee
The bound states for $n\ge1$ are doubly degenerate,
$\epsilon_{n,j}=\epsilon_{n,-j}$. The lowest energy level is given by
\be
\epsilon_{0,j=1/2} = m\sqrt{1-(2Z\A)^2}\,. \label{lowest}
\ee
If $Z\A$ exceeds $1/2$, then the energy (\ref{lowest}) becomes imaginary, i.e., the fall into the
center phenomenon \cite{Shytov1,Shytov2,Castro} occurs. According to
\cite{Pomeranchuk,Popov}, nonzero $R$ resolves this problem.
 For $Z\A>1/2$, $\nu$  is imaginary for certain $j$ and
for such $j$ we denote $\nu=i\beta,\beta=\sqrt{Z^2\alpha^2-j^2}$. For finite
$R$ discrete levels exist for $Z\alpha>1/2$. Their energy decreases with
increasing of $Z\A$ until they reach the lower continuum. The behavior of
lowest energy levels with $j=1/2$ as functions of the coupling $Z\alpha$ is
shown in Fig.\ref{cr1}.

The critical charge $Z_{c}$ that corresponds to diving into the continuum is
obtained from Eq.(\ref{eq:Rtozero2nd}) setting $\e = -m$ there and using the
corollary of the Stirling formula: $\frac{\Gamma(x+iy)}{\Gamma(x-iy)}
\rightarrow e^{2iy \log x} ,\,\,\,\, x \rightarrow +\infty$. We come at the
equation
\be
\label{crc}
e^{-2i\beta\log(2Z\A mR)} = \frac{i\beta - j +
Z\alpha\sigma(Z\alpha,j)}{-i\beta - j +
Z\alpha\sigma(Z\alpha,j)}\frac{\Gamma(1-2i\beta)}{\Gamma(1+2i\beta)}\,,
\ee
or,
\be -\beta\log(2Z\A mR)=\arg\,(Z\alpha\sigma({Z\alpha,j})-j+i\beta)
+\arg\,\Gamma(1-2i\beta)+\pi n, \label{critcoupling:eq}
\ee
where $n$ is integer. It is not difficult to check that for $j=1/2$ and $n=1$ the critical
coupling $Z_c\alpha$ approaches the value $1/2$ for $mR \rightarrow 0$. The
dependence of the critical coupling $Z_c\alpha$ on $mR$ for $j=1/2$ is shown in
Fig.\ref{c1}.

The bound and  quasistationary states in gapped graphene in the case of the supercritical
Coulomb impurity  were also numerically calculated
in the tight-binding lattice model which has a natural lattice scale cutoff that
provides an important control of the validity of the Dirac equation approach \cite{Pereira,Zhu}.

\section{Fourth order differential equation}
\label{Pochhammer:eq}
The system of equations (\ref{coupled-system})
with boundary conditions (\ref{boundary-conditions}) is reduced to
the following fourth order differential equation for the function
$\chi_{5}(q)$:
\be
\chi^{IV}_{5}+
\frac{2(5q^2-3p_0^2)}{q(q^2-p_0^2)}\chi^{'''}_{5} +
\frac{(44+5\lambda)q^2-8p_0^2}{2q^2(q^2-p_0^2)}\chi_{5}''+
\frac{4p_0^2+(8+7\lambda)q^2}{q^3(q^2-p_0^2)}\chi_{5}'+\frac{3\lambda^2\chi_{5}}{2q^2(q^2-p_0^2)}=0,
\ee
with the corresponding boundary conditions
\ba
&&q^{2}\chi'_5\Big|_{q=0}=0, \quad
q^{4}\frac{d}{dq}\left[(q^2-p_0^2)\left(\chi_{5}''+\frac{2}{q}\chi_{5}'
+\lambda \frac{\chi_{5}}{q^2-p_0^2}\right)\right]\Big|_{q=0}=0,\\
&& (q\chi_5(q))^{\prime}\Big|_{q=\Lambda}=0,\quad
 \frac{d}{dq}\left[q^3(q^2-p_0^2)\left(\chi_{5}''+\frac{2}{q}\chi_{5}'+\lambda
 \frac{\chi_{5}}{q^2-p_0^2}\right)\right]\Big|_{q=\Lambda}=0.
\ea
In terms of the variable $z=q^{2}/p^{2}_{0}$ these equations are
rewritten as
\be
\left[z^3(z-1)\frac{d^{4}}{dz^{4}}+2z^2(4z-3)\frac{d^{3}}{dz^{3}}+
\frac{5}{8}z((\lambda+22)z-10)\frac{d^{2}}{dz^{2}}+\frac{19\lambda+60}{16}z\frac{d}{dz}
+ \frac{3\lambda^2}{32}\right]\chi_5=0,
\label{higher-hypergeometric}
\ee
and boundary conditions
\ba
&&z^{3/2}\frac{d\chi_{5}}{dz}\Big|_{z=0}=0,\quad
z^{5/2}\frac{d}{dz}\left[
(z-1)\left(4z\frac{d^{2}\chi_{5}}{dz^{2}}+6\frac{d\chi_{5}}{dz}+\frac{\lambda\chi_{5}}
{z-1}\right)\right]\Big|_{z=0}=0,\\
&&
\left(2z\frac{d\chi_5}{dz}+\chi_5\right)\Big|_{z=\Lambda^2}=0,\quad
z^{1/2}\frac{d}{dz}\left[z^{3/2}(z-1)\left(4z\frac{d^2\chi_5}{dz^2}+
6\frac{d\chi_5}{dz}+\frac{\lambda
\chi_5}{z-1}\right)\right]\Big|_{z=\Lambda^2}=0.
\label{boundary-conditions2}
\ea
Eq.(\ref{higher-hypergeometric}) is
the Pochhammer-type equation \cite{Bateman1}, its canonical form is
\be
\left(\prod_{k=0}^{3}(\theta + b_k-1) - z
\prod_{k=1}^{4}(\theta+a_k)\right)\chi_5 = 0, \quad \theta\equiv
z\frac{d}{dz},
\ee
where the parameters
\ba
&& b_0=1,\quad b_1=3/2,\quad b_2=3/2,\quad b_3=0,\\
&&a_{1,2} = \frac{1}{4}\left(1 \pm \gamma\right),\quad
\gamma=\sqrt{1-4\lambda},\quad a_{3,4} =
\frac{3}{4}\left(1\pm\tilde{\gamma}\right),\quad
\tilde{\gamma}=\sqrt{1-\frac{2\lambda}{3}}
\ea
describe the behavior of $\chi_{5}(z)$ at the points $z=0$ and $z=\infty$,
respectively. The general solution of
Eq.(\ref{higher-hypergeometric}) at the point $z=0$ can be written in terms of
four linearly independent solutions,
\ba
\chi_5 &=&
\frac{C_1}{\sqrt{z}}\,_4F_3\left(a_1-\frac{1}{2},a_2-\frac{1}{2},a_3-
\frac{1}{2},a_4-\frac{1}{2}; -\frac{1}{2},\frac{1}{2},1;z\right) +
C_2\,z\,{}_4F_3\left(a_1+1,a_2+1,a_3+1,a_4+1; \frac{5}{2},\frac{5}{2},2;z\right)\nonumber\\
&+&
C_3 G^{24}_{44}\left(z\Big|\begin{array}{cccc}1-a_1,&1-a_2,&1-a_3,&1-a_4\\
-1/2,&-1/2,&0,&1\end{array}\right)
+ C_4 G^{34}_{44}\left(-z\Big|\begin{array}{cccc}1-a_1,&1-a_2,&1-a_3,&1-a_4\\
-1/2&0,&1,&-1/2\end{array} \right),
\ea
where ${}_{q+1}F_{q}((a)_{q+1};(b)_{q};z)$ is higher hypergeometric function
and
\ba
G^{mn}_{pq}\left(z\Big|\begin{array}{cccc}a_{1},&\dots a_{n},&a_{n+1},&\dots a_{p}\\
b_{1},&\dots b_{m},&b_{m+1},&\dots b_{q}\end{array}\right)
\ea
is the Meijer G-function \cite{Prudnikov}. Leading asymptotic of the
each term at $z \to 0$ is
\ba
\chi_5 &=& \frac{C_1}{\sqrt{z}}(1+O(z))+
zC_2(1+O(z))+ C_3 \frac{1}{\sqrt{z}}\frac{1}{2\pi}\prod\limits_{i=1}^{4}
\Gamma(a_{i}-1/2)\left(\log z+D+O(z)\right)\nonumber\\
 &+& C_4\left(-\frac{i\pi}{2\sqrt{z}}\prod\limits_{i=1}^{4}
\Gamma(a_{i}-1/2)+O(1) \right),\quad
D=4\gamma-2+4\log2+\sum\limits_{i=1}^{4}\psi(a_{i}-1/2),
\ea
and $\gamma$ is the Euler constant.
Hence from the boundary conditions (\ref{boundary-conditions2}) we
find that $C_3=0$ and
\be
C_{1}=C_{4}\frac{i\pi}{2}\prod\limits_{i=1}^{4}
\Gamma(a_{i}-1/2).
\ee
Asymptotical behavior of the function ${}_{4}F_{3}$ can be found from Eq.7.2.3.77
in the book \cite{Prudnikov}, thus we obtain
\be
{}_{4}F_{3}\left(a_{1},a_{2},a_{3},a_{4};b_{1},b_{2},b_{3};z\right)\simeq
\frac{\prod\limits_{i=1}^{3}\Gamma(b_{i})}{\prod\limits_{i=1}^{4}\Gamma(a_{i})}\sum\limits_{k=1}^{4}
(-z)^{-a_{k}}\frac{\Gamma(a_{k}){\prod\limits_{i=1}^{4}}\Gamma(a_{i}^{\prime}-a_{k})}
{\prod\limits_{i=1}^{3}\Gamma(b_{i}-a_{k})},
\ee
if no two $a_{k}, \, k=1,\dots 4$, differ by an integer,
the prime  in the product ${\prod\limits_{i=1}^{4}}\Gamma(a_{i}^{\prime}-a_{k})$
means that the term with $i=k$ is absent.
 Thus we obtain
\ba
&&z\,{}_4F_3\left(a_1+1,a_2+1,a_3+1,a_4+1; \frac{5}{2},\frac{5}{2},2;z\right)\simeq
-\frac{\Gamma^{2}(5/2)}{\prod\limits_{i=1}^{4}\Gamma(a_{i}+1)}\nonumber\\
&&\times\left[(-z)^{-a_{4}}
\frac{\Gamma(a_{4}+1)\Gamma(a_{1}-a_{4})\Gamma(a_{2}-a_{4})\Gamma(a_{3}-a_{4})}
{\Gamma^{2}(\frac{3}{2}-a_{4})\Gamma(1-a_{4})}+(\mbox{3 cyclic permutations}\,
1\to2\to3\to4\to1)\right].
\ea
Similarly, for the Meijer G-function we use Eq.8.2.1.4 in the book \cite{Prudnikov}
to find the asymptotic at large $z$:
\ba
G^{mn}_{pq}\left(z\Big|\begin{array}{cccc}a_{1},&\dots a_{n};&a_{n+1},&\dots a_{p}\\
b_{1},&\dots b_{m},&b_{m+1},&\dots b_{q}\end{array}\right)\simeq
\sum\limits_{k=1}^{n}z^{a_{k}-1}\frac{\prod\limits_{i=1}^{n}\Gamma(a_{k}-a_{i}^{\prime})
\prod\limits_{i=1}^{m}\Gamma(1+b_{i}-a_{k})}{\prod\limits_{j=m+1}^{q}\Gamma(a_{k}-b_{j})
\prod\limits_{j=n+1}^{p}\Gamma(1+a_{j}-a_{k})},
\ea
if no two $a_{k}, \, k=1,\dots n$, differ by an integer.

In our case we obtain
\ba
G^{34}_{44}\left(-z\Big|\begin{array}{cccc}1-a_1,&1-a_2,&1-a_3,&1-a_4\\
-1/2&0,&1,&-1/2\end{array} \right)&\simeq& (-z)^{-a_4} \frac{\Gamma(a_1-a_4)
\Gamma(a_2-a_4)\Gamma(a_3-a_4)\Gamma(a_4)\Gamma(1+a_4)\Gamma(a_4-1/2)}{\Gamma(3/2-a_4)}\nonumber\\
&+&(\mbox{3 cyclic permutations}\,
1\to2\to3\to4\to1).
\ea
Hence the function $\chi_{5}(z)$ behaves at $z\to\infty$ as
\ba
\chi_5(z) = \sum\limits_{i=1}^{4}A_{i}z^{-a_i}\left(1+O(1/z)\right),\quad
A_{i} = (-1)^{-a_i}\left({-C_4}\pi^{2}\cot(\pi a_{i})-C_2
\frac{\Gamma^{2}(5/2)}{\prod\limits_{i=1}^{4}\Gamma(a_{i}+1)}\right) F_{i},
\ea
where
\ba
F_{i}= \frac{\Gamma(a_j-a_i)\Gamma(a_k-a_i)\Gamma(a_l-a_i)\Gamma(1+a_i)}{\Gamma^{2}(3/2-a_i)
\Gamma(1-a_i)},\quad k\neq l\neq j \neq i.
\ea
The UV boundary conditions lead to the following equations:
\ba
&&A_1(1-\gamma)z^{-\gamma/4} +A_2(1+\gamma)z^{\gamma/4} - A_3 (1+3\tilde{\gamma})z^{-1/2-3\tilde{\gamma}/4}
-  A_4 (1-3\tilde{\gamma})z^{-1/2+3\tilde{\gamma}/4}=0,\\
&& A_1(1-\gamma)z^{-\gamma/4} + A_2(1+\gamma)z^{\gamma/4} + A_3\frac{3}{2} (3+\tilde{\gamma})
z^{1/2-3\tilde{\gamma}/4}+ A_4 \frac{3}{2}(3-\tilde{\gamma})z^{1/2+3\tilde{\gamma}/4} =0,
\ea
where $z=\Lambda^{2}/p^{2}_{0}$. This system of equations does not have solutions
for $\lambda<1/4$. Near the critical value, $\lambda\gtrsim1/4$, we can neglect
the terms with $A_{3}$, then we get $A_{4}=0$ and
\be
A_{1}(1-i\omega)z^{-i\omega/4}+A_{2}(1+i\omega)z^{i\omega/4}=0, \quad \omega=-i\gamma.
\ee
This gives
\be
C_{2}\frac{\Gamma^{2}(5/2)}{\prod\limits_{i=1}^{4}\Gamma(a_{i}+1)}=-C_{4}\pi^{2}\cot(a_{4}),
\ee
and the equation
\be
\left(-\frac{\Lambda^{2}}{p^{2}_{0}}\right)^{i\omega/2}=-\frac{1-i\omega}{1+i\omega}
\frac{F_{1}}{F_{2}}\frac{\cot(\pi a_{1})-\cot(\pi a_{4})}{\cot(\pi a_{2})-\cot(\pi a_{4})}.
\ee
From this equation we get when $\omega\ll1$,
\be
p^{2}_{0}=-\Lambda^{2}\exp[-\frac{4\pi n}{\omega}+a],\quad a\approx7.14,\quad n=1,2,\dots.
\ee
Comparing this result  with Eq.(\ref{tachyon}) we see that the approximation that
decouples the system (\ref{coupled-system}) works nicely near the critical coupling $\lambda_{c}$.

\end{document}